\documentclass[pdflatex,sn-mathphys-ay]{sn-jnl}

\usepackage{graphicx}%
\usepackage{multirow}%
\usepackage{amsmath,amssymb,amsfonts}%
\usepackage{amsthm}%
\usepackage{mathrsfs}%
\usepackage[title]{appendix}%
\usepackage{xcolor}%
\usepackage{textcomp}%
\usepackage{manyfoot}%
\usepackage{booktabs}%
\usepackage{algorithm}%
\usepackage{algorithmicx}%
\usepackage{algpseudocode}%
\usepackage{listings}%
\usepackage{bm}%
\usepackage{subcaption}%
\usepackage{gensymb}%

\theoremstyle{thmstyleone}%
\theoremstyle{thmstyletwo}%

\theoremstyle{thmstylethree}%

\begin{document}

\title{Discrete Gaussian Vector Fields on Meshes}


\author*[1]{\fnm{Michael} \sur{Gillan}}\email{m.gillan@exeter.ac.uk}

\author[1]{\fnm{Stefan} \sur{Siegert}}\email{s.siegert@exeter.ac.uk}

\author[1]{\fnm{Ben} \sur{Youngman}}\email{b.youngman@exeter.ac.uk}

\affil[1]{\orgdiv{Department of Mathematics and Statistics}, \orgname{University of Exeter}, \orgaddress{\city{Exeter}, \country{UK}}}

\abstract{Though the underlying fields associated with vector-valued environmental data are continuous, observations themselves are discrete. For example, climate models typically output grid-based representations of wind fields or ocean currents, and these are often downscaled to a discrete set of points. By treating the area of interest as a two-dimensional manifold that can be represented as a triangular mesh and embedded in Euclidean space, this work shows that discrete intrinsic Gaussian processes for vector-valued data can be developed from discrete differential operators defined with respect to a mesh. These Gaussian processes account for the geometry and curvature of the manifold whilst also providing a flexible and practical formulation that can be readily applied to any two-dimensional mesh. We show that these models can capture harmonic flows, incorporate boundary conditions, and model non-stationary data. Finally, we apply these models to downscaling stationary and non-stationary gridded wind data on the globe, and to inference of ocean currents from sparse observations in bounded domains.}


\maketitle

\clearpage

\section{Introduction}\label{sec1}

Statistical downscaling of climate models focuses on the statistical relationship between spatial locations and dynamical models. High-quality inference procedures and the availability of historic weather data allow us to make inferences about what the model outputs would be at unobserved spatial locations. This is useful when the outputs of climate models are fed into other dynamical models.

The goal of downscaling is to take observations of some variable(s) of interest at some spatial locations $s_1, \dots, s_n$ and infer the value of the variable(s) at some unobserved locations $s_{n+1},\dots, s_m$. Formulated in this way it is a regression problem, and a common methodology is to use Gaussian processes (GPs). A GP can loosely be considered to be a distribution over functions, and more concretely be considered a multivariate distribution where the joint distribution of every set of variables is a Gaussian distribution \citep{rasmussenGaussianProcessesMachine2006}. When considering random fields, a GP becomes a prior over functions of space, where each unobserved spatial index has a mean and variance derived from the covariance kernel of the GP and observed data points. There are many examples of the application of this methodology to the problem of statistical downscaling in the literature; see \cite{xiongDataFusionGaussian2021} and \cite{fuentesModelEvaluationSpatial2005}. 

This work primarily considers the downscaling of vector-valued fields, defined as functions $F: \Omega \rightarrow \mathbb{R}^d$ for some input space $\Omega$ and some output dimension $d$. An obvious choice here would be to model the $u$ and $v$ components of the vector with independent scalar GPs, but this approach has limitations. In particular, when considering meteorological data, it leads to inconsistent secondary measurements, such as vorticity and divergence. A more flexible approach for capturing physical properties of the system, demonstrated by \cite{berlinghieriGaussianProcessesHelm2023}, is to use the differential and additive properties of GPs to develop a covariance kernel based on the Helmholtz-Hodge decomposition (HHD) \citep{bhatiaHelmholtzHodgeDecompositionSurvey2013}. The HHD decomposes a vector-valued field into a curl-free field, a divergence-free field, and a curl- and divergence-free (harmonic) field. \cite{berlinghieriGaussianProcessesHelm2023} establish a univariate spatial GP prior on the decomposed components, which allows them to create a closed-form prior on the reconstructed field. However, this model has limited applications on non-flat domains, such as the sphere, and omits harmonic fields. 

The use of GPs to model quantities on manifolds is an area of significant recent study. \cite{borovitskiyMaternGaussianProcesses2021} develops a theory of scalar \textit{intrinsic} GPs in manifold settings through the functional calculus of the Laplacian operator. \cite{borovitskiyMaternGaussianProcesses2023} extends this to a discretised setting by replacing the smooth Laplacian with the graph Laplacian. Both \cite{hutchinsonVectorvaluedGaussianProcesses2021} and \cite{robert-nicoudIntrinsicGaussianVector2024} present methods for inference on vector fields in manifold settings and apply them to wind field data. The former models 10m wind speed and direction, which can exhibit significant non-stationarity and anisotropy, whereas the latter models high-altitude geostrophic winds using a divergence-free kernel. \cite{wyrwalResidualDeepGaussian2024} introduce deep GP formulations of the above models and apply them to non-stationary, lower-altitude wind fields. \cite{yangHodgeCompositionalEdgeGaussian2024} proposes a methodology for modelling functions on the edges of meshes, extending the formulation of intrinsic vector GPs with the graph Laplacian. They apply this to interpolating ocean currents in the Pacific. 

In this work, we show that the methodology of \cite{robert-nicoudIntrinsicGaussianVector2024} has an exact discrete analogue by presenting a formulation of discrete intrinsic vector GPs through the functional calculus of the cotangent Laplacian and discrete exterior calculus operators. As in the frameworks above, this model can capture the different components of the HHD and model vector fields on manifolds. Unlike previous works, we show that this methodology is flexible enough to model vector fields on arbitrary discretised 2-dimensional manifolds, and easily extends to incorporate boundary conditions and non-stationarity. We demonstrate the ability of this methodology on downscaling wind fields and ocean currents in a variety of domains.

\section{Background} \label{sec:bg}

\subsection{Intrinsic Gaussian processes} \label{sec:intrinsic-gps}

\cite{whittleStochasticProcessesSeveral1963} formalises the Matérn GP as the solution to the SPDE 
\[
(\kappa^2 - \Delta)^{\alpha/2}x(\mathbf{s}) = \mathcal{W}(\mathbf{s}),
\]
for length-scale and smoothness parameters $\kappa$ and $\alpha$, and a white noise process $\mathcal{W}$. This result is used by \cite{lindgrenExplicitLinkGaussian2011} to explicitly construct Matérn precision matrices with respect to a mesh by discretising the SPDE using piece-wise linear basis functions. As highlighted in a later review \citep{lindgrenSPDEApproachGaussian2022}, this result can be generalised to smooth manifolds by replacing the Euclidean $\Delta$ with the equivalent manifold operator $\Delta_g$. The SPDE approach is a pioneering work, but does not readily extend to modelling vector-valued fields.

\cite{borovitskiyMaternGaussianProcesses2023} introduce an alternative formulation for scalar covariance kernels defined on an arbitrary Riemannian manifold $(M, g)$ through the spectral decomposition of the Laplace-Beltrami operator $-\Delta_g$. For eigenvalues $\lambda_i$ and eigenfunctions $f_i$, they show that  Matérn and squared exponential GPs in the sense of \cite{whittleStochasticProcessesSeveral1963} and \cite{lindgrenExplicitLinkGaussian2011} can be defined as 

\begin{align*}
k_\nu(x, x') &= \frac{\sigma^2}{C_\nu} \sum_{n=0}^\infty \left(\frac{2\nu}{\kappa^2} + \lambda_n\right)^{-\nu - \frac{d}{2}} f_n(x) f_n(x'), \\
k_\infty(x, x') &= \frac{\sigma^2}{C_\infty} \sum_{n=0}^\infty e^{-\frac{\kappa^2}{2} \lambda_n} f_n(x) f_n(x'),
\end{align*}
with normalising constant $C_{(\cdot)}$ and hyperparameters $\kappa$, $\nu$ and $\sigma$, controlling length-scale, smoothness, and variance, respectively. The sum to $\infty$ above gives the ``true" kernel, but \cite{robert-nicoudIntrinsicGaussianVector2024} highlight that any truncation of this sum (i.e. to $L$ terms) gives a well-defined, albeit lower rank, covariance kernel. \cite{borovitskiyMaternGaussianProcesses2021} show that the above formulation can be applied to graphs and triangular meshes through the eigenpairs of the graph Laplacian, and show empirically that it converges to its smooth counterpart in the Riemannian setting. For simplicity, we define a function $\Phi$ for the eigenvalue scaling in Equation \ref{eq:scale},

\begin{equation}
\Phi_{\nu, \kappa}(\lambda) =
\begin{cases}
\left(\frac{2\nu}{\kappa^2} + \lambda\right)^{-\nu - \frac{d}{2}}, & \nu < \infty, \\
e^{-\frac{\kappa^2}{2} \lambda}, & \nu = \infty.
\end{cases}
\label{eq:scale}
\end{equation}

\cite{robert-nicoudIntrinsicGaussianVector2024} extend the ideas in \cite{borovitskiyMaternGaussianProcesses2023} to vector fields on $M$. They provide the formulation

\begin{equation*}
    k_{\nu, \kappa, \sigma^2}(x, x') \propto \sigma^2 \sum_{n=0}^L \Phi_{\nu, \kappa}(\lambda_n) s_n(x) \otimes s_n(x'),
\end{equation*}
where $s_n$ are the eigenfunctions of the Hodge Laplacian on a two-dimensional manifold; see \cite{rosenbergLaplacianRiemannianManifold1997a}. More appropriately for this work, they provide an alternative formulation, \textit{Hodge-compositional Matérn kernels}, based on the HHD. By finding eigenfunctions of the Hodge Laplacian that are divergence-free ($s_i^{(curl)} = \nabla f$), curl-free ($s_i^{(div)} = \ast \nabla f$), and harmonic ($\Delta s_i^{(harm)} = 0)$, kernels for each of these classes can be constructed independently. This allows parameterisation based on curling and diverging velocities, and accounts for instances where the data may be a vector field with only one of these parts. Eigenfunctions $s_n^{(curl)}$ and $s_n^{(div)}$ can be found by finding the gradient and curl of the eigenfunctions $f_i$ of the Laplace-Beltrami operator, giving curl-free and divergence-free vector fields respectively. $s_n^{(harm)}$ are an orthonormal basis of constant vector fields in the domain, and can be generally found as the 0-eigenspace of the Hodge-Laplacian. \\

\subsection{Discrete Exterior Calculus} \label{sec:dec}

Exterior calculus is the calculus defined by generalising the derivative to operate on higher-degree differential forms. A differential $k$-form is a quantity that can be integrated over a k-dimensional manifold, e.g. the differential of a scalar function $g(x) dx$ is a 1-form and can be integrated over a 1-dimensional line segment by $\int_a^b g(x) dx$. 1-forms are the natural dual of vector fields, and vector calculus identities expressed in exterior calculus are metric independent, meaning they hold on any differentiable manifold. 

The language of differential forms provides an obvious path for discretisation: a smooth 2-dimensional Riemannian manifold $M$ can be discretised as a collection of connected 0-, 1-, and 2- dimensional piecewise linear components. These are, respectively, vertices, edges, and faces, and the collection is referred to as a simplicial complex (or mesh). k-forms can be integrated over this mesh as they would the smooth manifold, thus providing an exact discrete analogue to exterior calculus, referred to as Discrete Exterior Calculus (DEC) \citep{hiraniDiscreteExteriorCalculus2003}. An implementation of most of the core theory is provided in the PyDEC library \citep{bellPyDECSoftwareAlgorithms2012}. The mesh is embedded in Euclidean space through an isomorphism, $\varphi$, which ensures the induced metric is equal to the Riemannian metric $g$ of the manifold $M$. Such an embedding exists via the $\mathcal{C}^1$ Nash-Kuiper embedding theorem \citep{kuiperC1isometricImbeddings1955}.

DEC defines discrete versions of differential operators through the exterior derivative $\mathbf{d}_k$ and the Hodge star $\star_k$, where the subscript indicates the degree of differential form to which it is applied. $\mathbf{d}_k$ acts as a discrete difference operator for elements of the mesh, and can be thought of as analogous to a derivative. $\star_k$ captures geometric information of the embedded mesh that allows inner products defined on the mesh to obey the metric $g$. These inner products therefore approximate integration on the manifold $M$. As the number of vertices of the mesh increases, the mesh \textit{refines} (the maximal edge length $h$ tends to 0) and the discrete exterior derivative and Hodge star converge to their smooth counterparts \citep{schulzConvergenceDiscreteExterior2020}.

Differentials in DEC are integrated as numerical values over the directed edges of the simplicial complex. As indicated previously, 1-forms are duals of vector fields, but an additional step is required to convert from integrated 1-forms (a scalar value attached to each edge) to vector fields (three values indicating direction at each vertex). This is the \textit{sharp} operator, $(\cdot)^\#$. There are several definitions of sharp operator based on the source and destination forms, and the one used in this work is defined by \cite{hiraniDiscreteExteriorCalculus2003} as the primal-primal sharp, since it takes values on primal edges and produces vectors on primal vertices. This is achieved through an angle-weighted interpolation of the edge values in the one-ring around each vertex.

\subsection{Cotangent Laplacian}

There are a number of different discrete Laplace-Beltrami operators in the literature \citep{wardetzkyDiscreteLaplaceOperators2008}. Here the cotangent Laplacian is the most appropriate because it exhibits the most desirable property, \textit{convergence}. \cite{hildebrandtConvergenceMetricGeometric2006} show that the cotangent Laplacian $\mathcal{L}_c$ converges to the Laplace-Beltrami operator, $\Delta_g$, as the mesh converges to $M$. 

\cite{Wardetzky2007} defines $\mathcal{L}_c := \mathcal{M}^{-1}\mathfrak{L}$, where $\mathcal{M}$ is the mass matrix that encodes the $L^2$ inner product and $\mathfrak{L}$ is the stiffness matrix that represents the \textit{conformal} Laplacian. The conformal Laplacian encodes the cotangent formula of \cite{pinkallComputingDiscreteMinimal1993} and can be constructed in the DEC setting as $\mathbf{d}_0^\top \star_1 \mathbf{d}_0$ \citep{craneDigitalGeometryProcessing2013}. From the previous section, the mass matrix $\mathcal{M}$ is $\star_0$. Note that this is equivalent to the standard definition of the Laplacian as the divergence of the gradient, where the gradient is encoded in the discrete exterior derivative $\mathbf{d_0}$, and the term $\star_0^{-1}\mathbf{d}^\top_0\star_1$ is the divergence. The spectral decomposition of the cotangent Laplacian is therefore given by solutions to the generalised eigenproblem $\mathfrak{L}f = \lambda \mathcal{M} f$.

\section{Methodology} \label{sec:kernels}

\subsection{Discrete Vector Kernels} \label{sec:VMK} 

The work discussed in Section \ref{sec:bg} provides the building blocks for vector-valued kernels on discretised manifolds. The first step is to construct a simplicial complex for the region of interest, ensuring that the observation and inference points lie on the vertices of the mesh. The cotangent Laplacian $\mathcal{L}_c := \star_0^{-1}\mathbf{d}_0^\top \star_1 \mathbf{d}_0$ can be constructed from DEC operators defined with respect to this mesh, and eigenpairs of $\mathcal{L}_c$ can be found by solving the generalised eigenproblem\footnote{For flat domains, we recommend replacing $\star_0$ with the identity. The metric in these domains is Euclidean, and the inclusion of the $\star_0$ simply introduces error as a result of boundary artifacts.} $\mathbf{d}_0^\top \star_1 \mathbf{d}_0f = \lambda \star_0 f$. The cotangent Laplacian $\mathcal{L}_c$ acts as the discrete equivalent of $\Delta_g$, and its eigenpairs provide a discrete basis for the definition of intrinsic GPs in the previous section. The eigenvectors derived in this way are orthonormal with respect to the vertex mass matrix, $f_i^\top \star_0 f_j = \delta_{ij}$. Subsequently, applying the exterior derivative to the eigenvectors makes them orthogonal with respect to $\star_1$, $(\mathbf{d}f_i)^\top \star_1 (\mathbf{d}f_j) = \lambda_i \delta_{ij}$, with a scaling required to ensure orthonormality, $(\frac{\mathbf{d}f_i}{\sqrt{\lambda_i}})^\top \star_1 (\frac{\mathbf{d}f_j}{\sqrt{\lambda_j}}) = \delta_{ij}$.

With the scaling in Equation \ref{eq:scale}, we can define a scalar covariance kernel on the vertices of the mesh,

\begin{equation}
    \mathbf{K}_{\nu, \kappa, \sigma^2} = \frac{\sigma^2}{C_{\nu, \kappa}} F \Phi_{\nu, \kappa}(\Lambda) \star_0 F^\top,
     \label{eq:scalar}
\end{equation}
where $F$ is the matrix formed by stacking the eigenvectors $f_i$ of $\mathcal{L}_c$. Note that capital, bold $\mathbf{K}$ is used to highlight that this is a covariance matrix, not a covariance function. such as $k$ in the previous section.
As in \cite{robert-nicoudIntrinsicGaussianVector2024}, $C_{\nu, \kappa}$ is a constant that ensures the variance of $\mathbf{K}$ is $\sigma^2$, given by $C_{\nu, \kappa} = \frac{1}{A} \sum_{n=0}^L \Phi_{\nu, \kappa}(\lambda_n)$. $A$ is the total area of the faces of the simplicial complex, which approximates the volume of the manifold $M$. 

Following \cite{robert-nicoudIntrinsicGaussianVector2024} we apply the discrete exterior derivative and Hodge star operators to each of the eigenvectors $f_n$ of $\mathcal{L}_c$ to form a basis for the Hodge-Laplacian. Given a harmonic basis for the surface (see Section \ref{sec:harm}), we can construct a covariance kernel as

\[\mathbf{K^d_{\nu, \kappa_d}} = \frac{1}{C_{\nu, \kappa_d}} \sum_{n=0}^L \Phi_{\nu, \kappa_d}(\lambda_n) f^d_n \otimes f^d_n,\]
\[\mathbf{K^c_{\nu, \kappa_c}} = \frac{1}{C_{\nu, \kappa_c}} \sum_{n=0}^L \Phi_{\nu, \kappa_c}(\lambda_n) f^c_n \otimes f^c_n,\]
\[\mathbf{K^h} = \sum_{n=0}^d f^h_n \otimes f^h_n,\] 
\[\mathbf{K^v} = \sigma_d^2 \mathbf{K^d} + \sigma_c^2 \mathbf{K^c} + \sigma_h^2 \mathbf{K^h},\]

where $f^d_n = (\frac{\mathbf{d} f_n}{\sqrt{\lambda_n}})^\#$, $f^c_n = \ast(\frac{\mathbf{d}f_n}{\sqrt{\lambda_n}})^\#$, $\{f^h_n\}_{0\leq n \leq d}$ is the harmonic basis of the domain, and $\otimes$ is the vector outer product. $\mathbf{K^d}$, $\mathbf{K^c}$, and $\mathbf{K^h}$ are covariance matrices for the curl-free, divergence-free, and harmonic vector fields, respectively. $L$ is chosen to capture sufficient local detail, usually set to 250 in practice, though it is more context-dependent than the smooth version since the maximum value of $L$ is the number of vertices in the mesh.

$\ast$ is used here to indicate that the application of the Hodge star to the diverging vector basis is performed via a positive $\frac{\pi}{2}$ rotation about the surface normal. On a 2-dimensional manifold, the application of the primal-primal sharp followed by this rotation is equivalent to application of the Hodge star followed by the application of a dual-primal sharp. To avoid the need to define a dual-primal sharp, we opt for the first approach. As with the primal-primal sharp, vertex normals are calculated via a angle-weighted sum of the incident face normals.

\begin{figure}
    \begin{subfigure}{.33\textwidth}
        \centering
        \includegraphics[width=\linewidth]{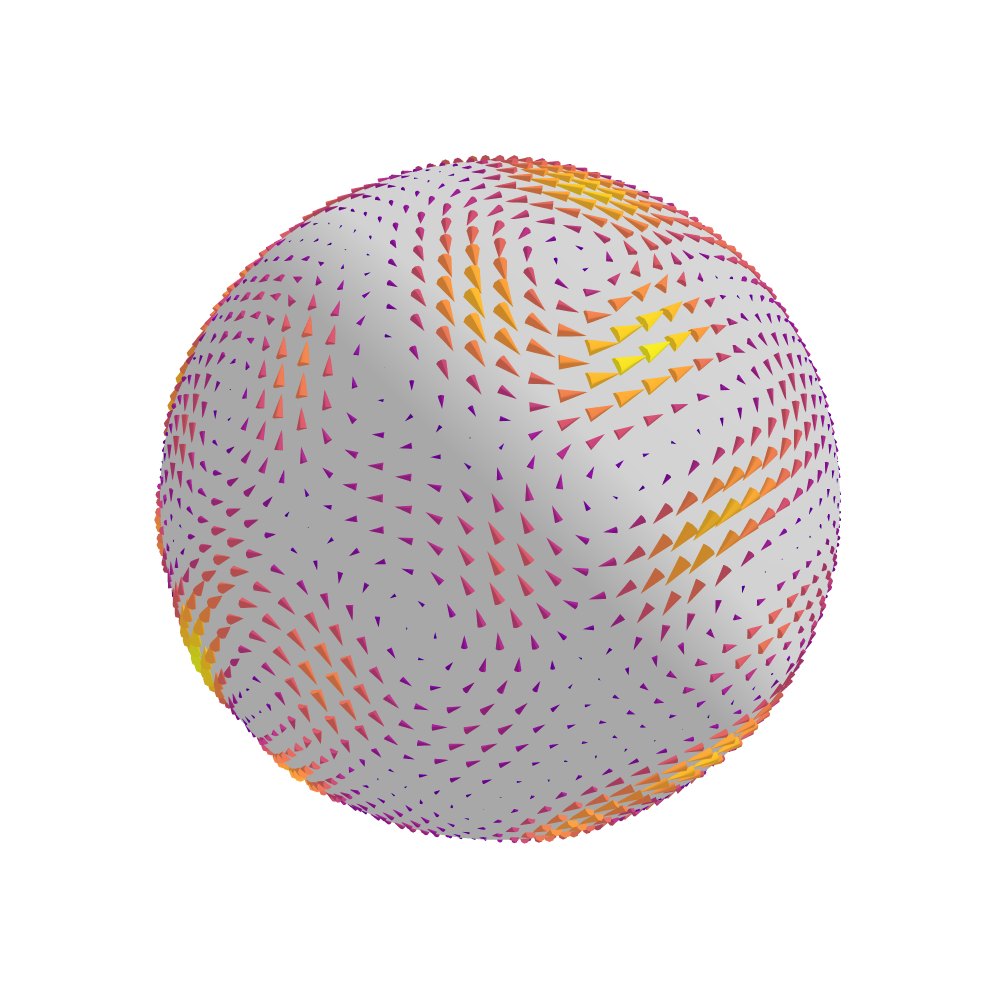}
        \label{fig:sph-mesh}
    \end{subfigure}%
    \begin{subfigure}{.33\textwidth}
        \centering
        \includegraphics[width=\linewidth]{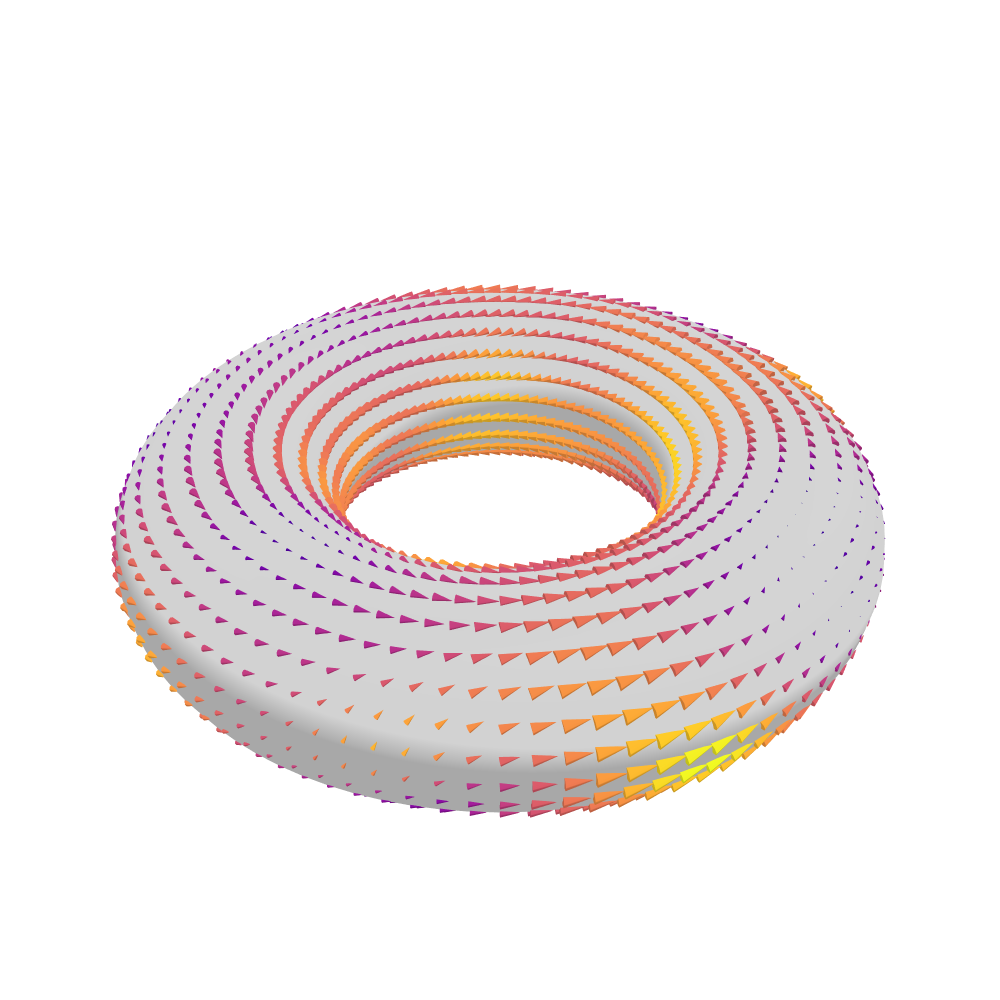}
        \label{fig:tor-mesh}
    \end{subfigure}%
    \begin{subfigure}{.33\textwidth}
        \centering
        \includegraphics[width=\linewidth]{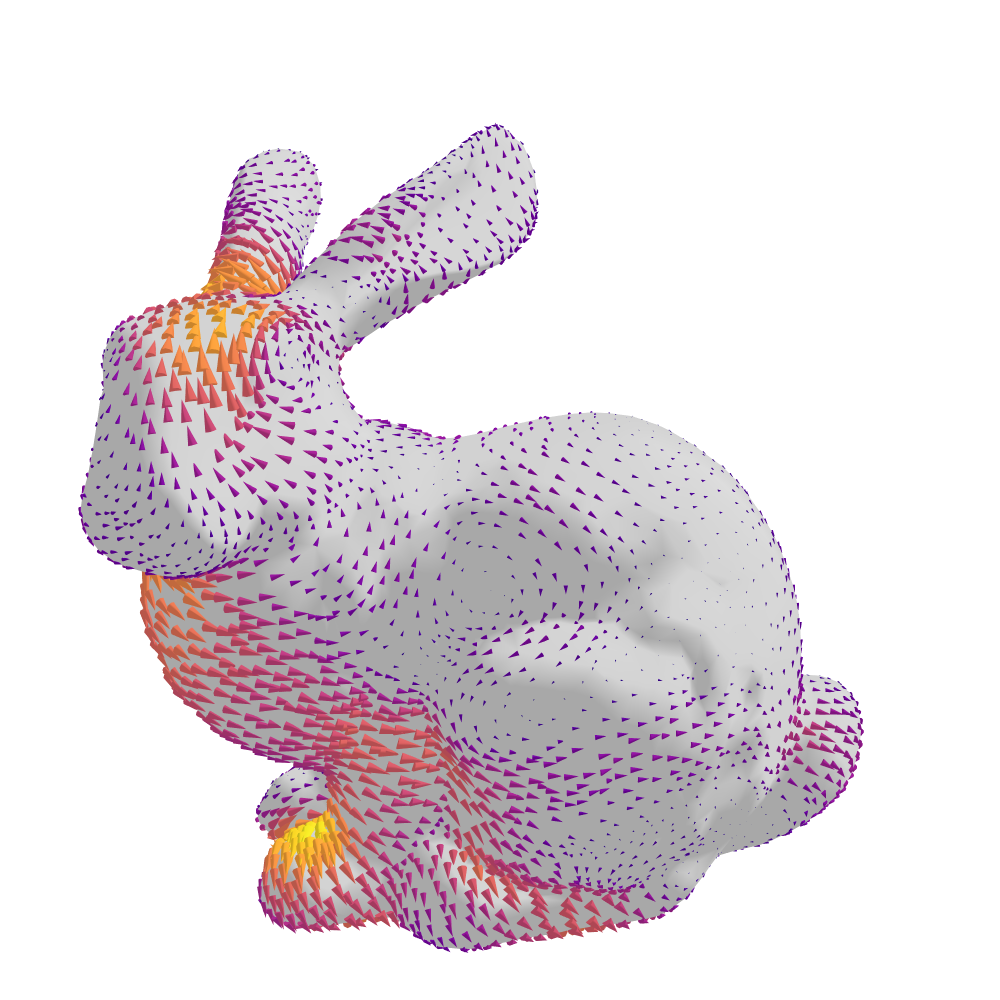}
        \label{fig:bun-mesh}
    \end{subfigure}
    \caption{Sampled vector fields on a variety of meshes. Colour and length of cones indicates magnitude. Left: an icosphere subdivided 4 times, giving 2562 vertices and 5210 faces. Middle: A torus with 96 major sections and 24 minor sections, giving 2304 vertices and 4608 faces. Right: Stanford Bunny, reduced to have 5048 vertices and 10000 faces.}
    \label{fig:meshes}
\end{figure}

This formulation of discrete vector kernels is incredibly flexible, generalising to arbitrary two-dimensional meshes with no adjustment. Figure \ref{fig:meshes} shows samples of discrete vector kernels on some interesting meshes.

\subsection{Harmonic Flows} \label{sec:harm}

\begin{figure}
    \centering
    \includegraphics[width=\linewidth]{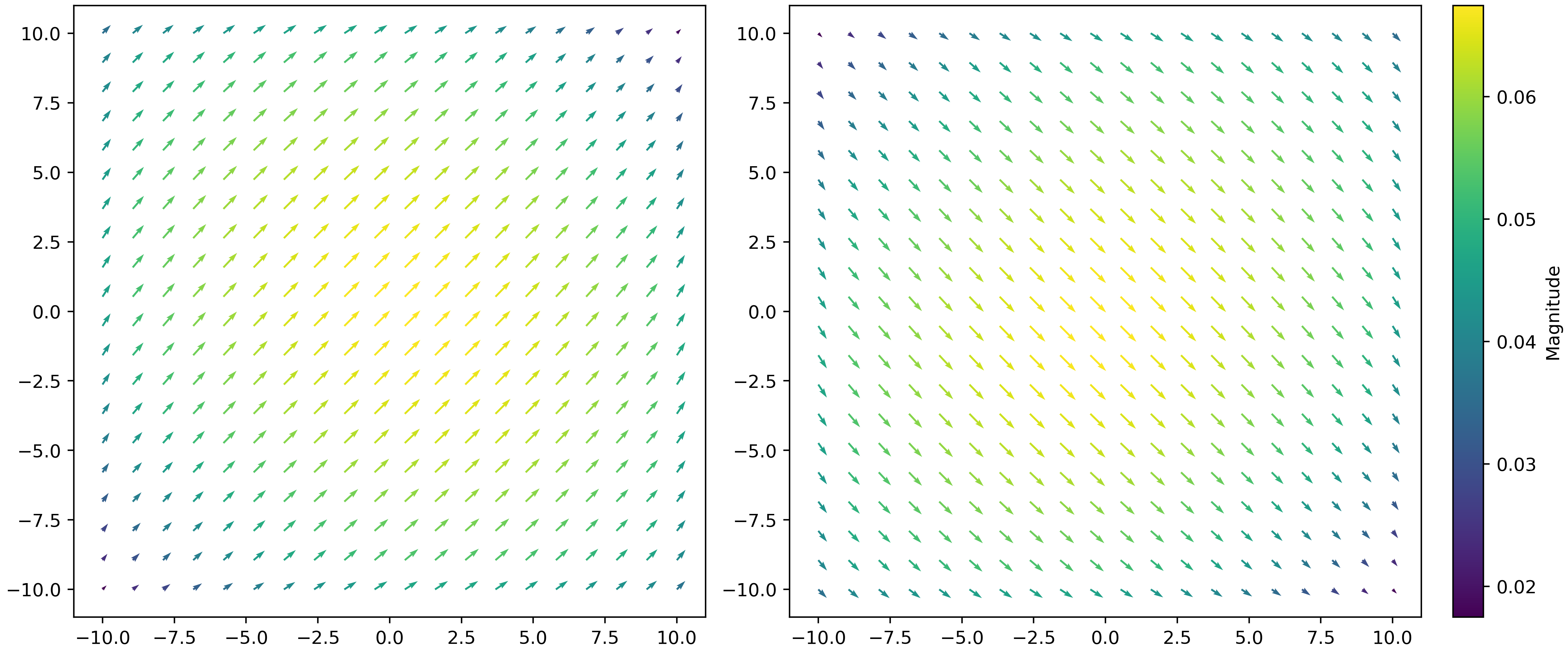}
    \caption{0-eigenspace of Hodge-Laplacian on a subset of $\mathbb{R}^2$. Colour indicates magnitude of the vector. There are obvious boundary effects due to the formulation of the Hodge-Laplacian as a DEC operator.}
    \label{fig:harm}
\end{figure}

In order to model harmonic components of flows, i.e. constant vector fields, we need a basis for vector fields in the null-space of the Hodge-Laplacian, which \cite{robert-nicoudIntrinsicGaussianVector2024} notes is the 0-eigenspace of the Hodge-Laplacian. A straightforward definition of the discrete Hodge-Laplacian applicable for this usage is given in \cite{bellPyDECSoftwareAlgorithms2012} as the weak form of the Laplace-de Rahm operator. Figure \ref{fig:harm} shows the 0-eigenspace of this operator over a subset of $\mathbb{R}^2$. The harmonic forms found in this way are orthonormal with respect to $\star_1$.

\begin{figure}
    \begin{subfigure}{.5\textwidth}
        \centering
        \includegraphics[width=\linewidth]{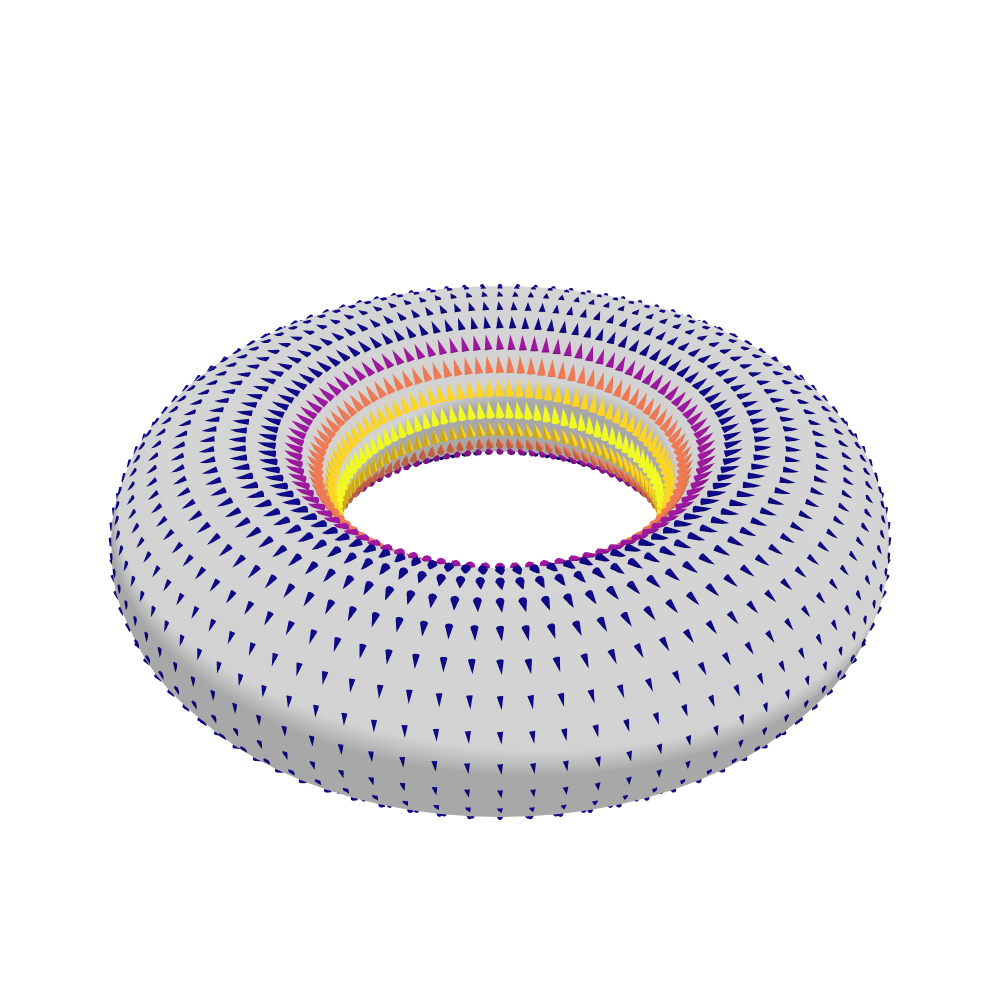}
    \end{subfigure}%
    \begin{subfigure}{.5\textwidth}
        \centering
        \includegraphics[width=\linewidth]{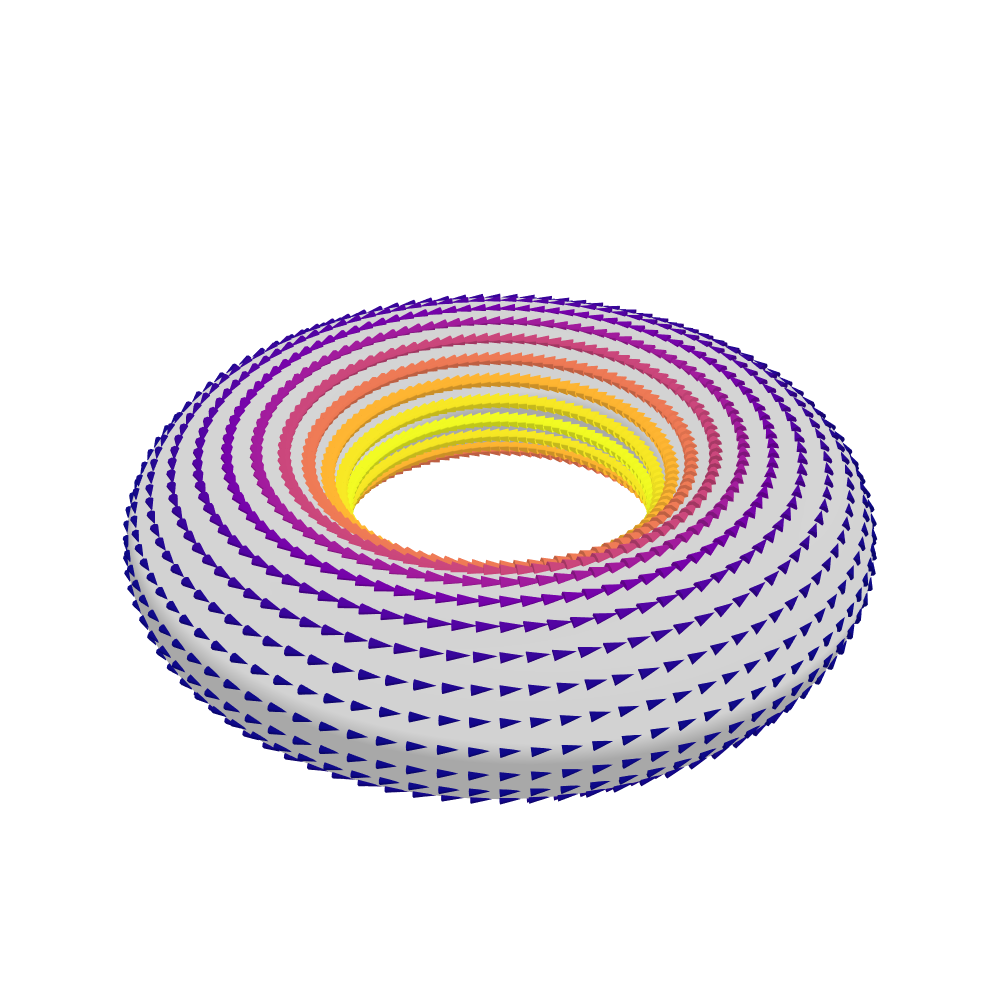}
    \end{subfigure}%
    \caption{The 0-eigenspace of the Hodge-Laplacian on a torus-shaped mesh. These two vector fields form an approximate harmonic basis on the torus. The increased magnitude of vectors on the interior of the torus is a product of the curvature introduced by the chosen ratio of major and minor radii, and all vector magnitudes approach 1 as the mesh approaches a flat torus.}
    \label{fig:torus-harmonics}
\end{figure}

Figure \ref{fig:harm} clearly shows that this formulation of the Hodge-Laplacian has natural Neumann boundary conditions when $\Omega$ has a boundary (i.e. $\Omega \subset \mathbb{R}^2$), and so this way of forming the harmonic basis only works on unbounded manifolds where the 0-eigenspace exists, such as a torus. Figure \ref{fig:torus-harmonics} shows the 0-eigenspace of a torus-like mesh. There are two obvious practical solutions for the other cases: we can define a harmonic basis ourselves, which is trivial for subsets of $\mathbb{R}^2$, or we can create an outer boundary that is far enough away from our area of interest that the impact of these boundary conditions is negligible; see \cite{lindgrenExplicitLinkGaussian2011}. 

\subsection{Boundary Conditions}

Boundary conditions can be imposed on the vector fields through conditions on the eigenvectors of the Laplace-Beltrami operator. In this section we show how a boundary condition of no flux can be enforced.

In order to create a 0-flux boundary condition, we must impose different boundary constraints on the diverging and curling basis fields. The condition for diverging fields is a Neumann boundary condition, which states that there is no flow in the direction of the outward normal $\hat{\mathbf{n}}$ at any point on the boundary $\partial\Omega$,

\[
\mathbf{d}f(s) \cdot\hat{\mathbf{n}}(s) = 0, \forall s \in \partial\Omega.
\]
This is naturally enforced by the discrete exterior derivative since there are no points beyond the boundary.

Recalling that a positive rotation is applied to the diverging fields to obtain the curling fields, the boundary condition for the curling fields is that, prior to this rotation, there is no flow tangential to the boundary. This can be achieved by requiring Dirichlet boundary conditions on the eigenvectors of the cotangent Laplacian:

\begin{alignat*}{2}
    \mathfrak{L}f &= \lambda \mathcal{M} f \quad &&\text{in } \Omega, \\
    f &= 0 \quad &&\text{on } \partial\Omega.
\end{alignat*}
This is similar to the eigenproblem posed in \citet[Section 11.1]{bellPyDECSoftwareAlgorithms2012}, and is solved in the same fashion: the boundary vertices are excluded from the eigenvalue problem, and the found eigenvectors are then augmented with 0s at the boundary vertices. The Neumann and Dirichlet eigenvectors can then be used as the basis for the diverging and curling kernels, respectively, giving a full vector kernel that obeys the no-flux boundary condition. It is worth noting however that as a result of the interpolation from 1-forms to vector fields, some small flux over the boundary may be introduced. This can be easily post-processed by removing the component of flow on the boundary in the direction of the normal.

\subsection{Non-Stationarity}

As highlighted in Section \ref{sec:intrinsic-gps}, this model is linked to the SPDE formulation of Matérn kernels presented in \cite{lindgrenExplicitLinkGaussian2011}. \cite{lindgrenSPDEApproachGaussian2022} modifies the differential operator such that the solutions are non-stationary Matérn fields:

\begin{equation*}
\{\kappa(\mathbf{s})^2 - \Delta \}^{\alpha/2} \tau(\mathbf{s}) u(\mathbf{s}) = \mathcal{W}(\mathbf{s)}.
\end{equation*}

This gives a non-stationary version of the functional calculus scaling in Equation \ref{eq:scale},

\begin{equation}
\Phi_{\nu, \kappa}(\lambda, \mathbf{s}) =
\begin{cases}
\left(\frac{2\nu}{\kappa(\mathbf{s})^2} + \lambda\right)^{-\nu - \frac{d}{2}}, & \nu < \infty, \\
e^{-\frac{\kappa(\mathbf{s})^2}{2} \lambda}, & \nu = \infty,
\end{cases}
\end{equation}
for some strictly positive function of space $\kappa(\mathbf{s})$. Whilst previous formulations introduce anisotropy at this level, we omit it for now in favour of a model that can be defined through the functional calculus of $\mathcal{L}_c$, as otherwise it would require that we perform the eigen-decomposition every time we wish to consider new anisotropy parameters.

The non-stationary scaling can be applied to the curling and diverging basis vector fields to produce a non-stationary covariance kernel. This kernel can then be used to infer non-stationary structure from data.

\section{Examples}

This section demonstrates how the methodology introduced in this work can be used to downscale vector-valued environmental data. First, it is applied to monthly-average, high-altitude wind fields across the entire globe, showing how the discrete methodology captures the metric of curved domains. We then infer non-stationary structure from hourly high-altitude wind fields. Finally, the methodology is applied in a bounded domain to downscale ocean currents. The code for these examples can be found at \url{https://github.com/michaelgillan1/DiscreteVectorFields}.

\subsection{Monthly Average High-Altitude Wind Fields} \label{sec:wind}

The first application of this methodology is to downscaling high-altitude wind field data across the globe. The data are the monthly averaged wind $u$ and $v$ components at the 500hPa pressure level for the month of July 2020, taken from the ERA5 atmospheric reanalysis dataset \citep{hersbachERA5MonthlyAveraged2023}. ERA5 outputs variables at 0.25\degree \  grid points, but due to computation limits we downscale from a 10\degree \ grid to a 2\degree \ grid. The data are normalised such that our observed data (the vectors at every 10th grid point) have an average norm of 1.

\begin{figure}[!ht]
    \centering
    \includegraphics[width=\linewidth]{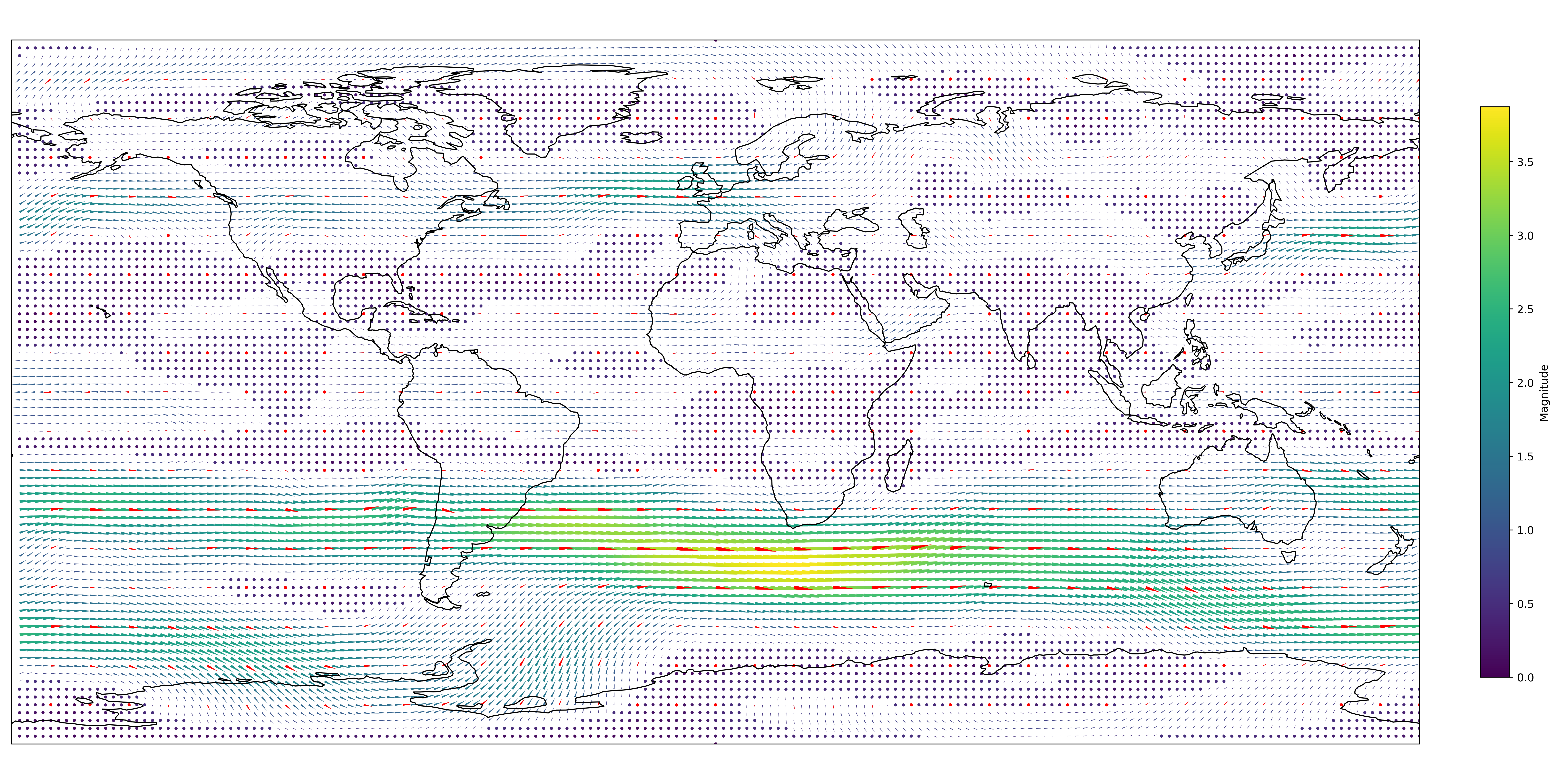}
    \caption{Downscaled monthly average wind data from ERA-5 on the plate carrée projection. Red arrows indicate observed data for the downscaling procedure, with size indicating magnitude. Other arrows are the posterior mean of the GP, with colour and size indicating magnitude. MSE = 0.027, NLL = -652.992}
    \label{fig:wind1}
\end{figure}

\begin{figure}[!ht]
    \centering
    \includegraphics[width=\linewidth]{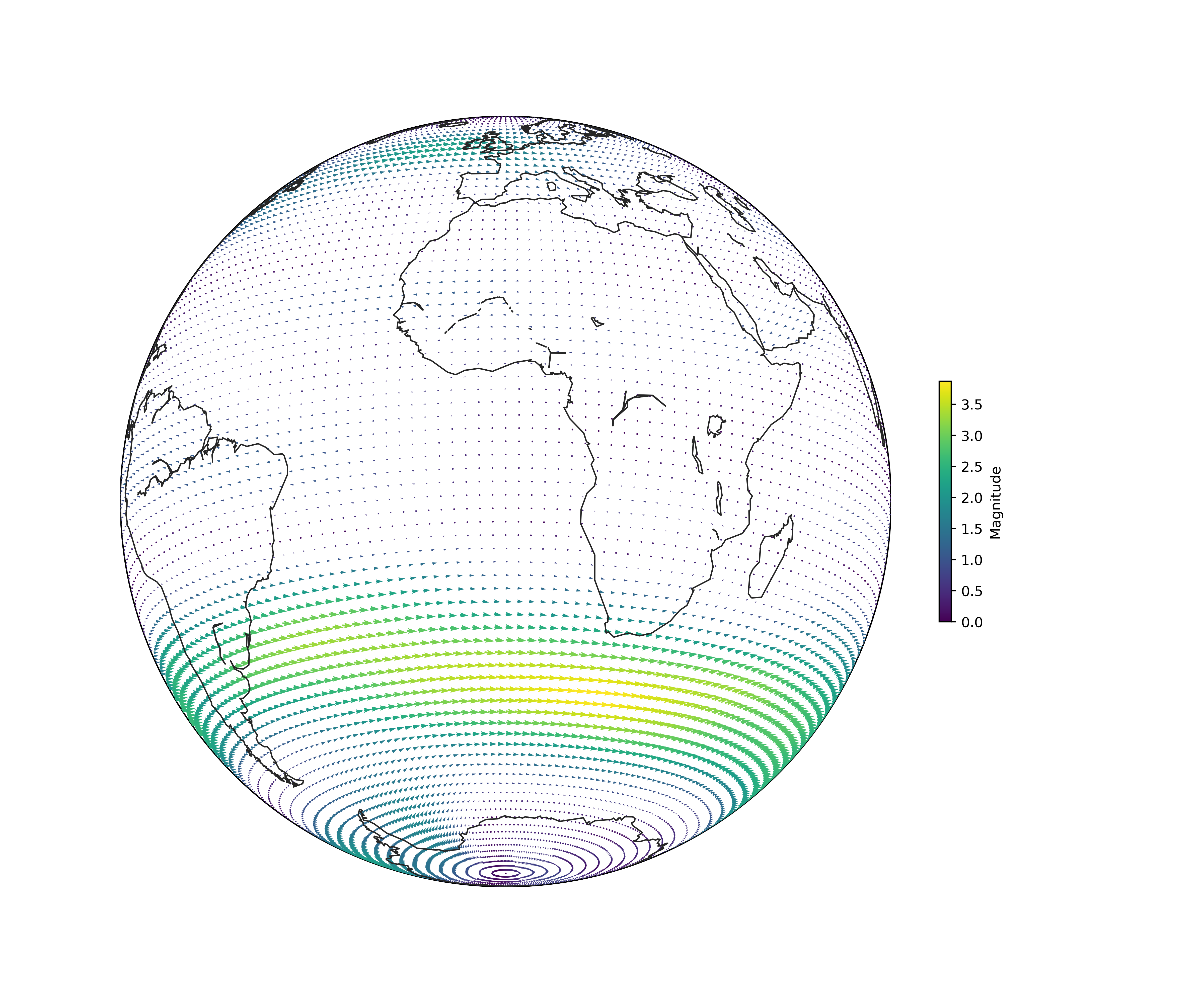}
    \caption{Downscaled monthly average wind data from ERA-5 on an orthographic projection. Arrow colour and size indicates magnitude.}
    \label{fig:wind2}
\end{figure}

\begin{figure}[!ht]
    \centering
    \includegraphics[width=\linewidth]{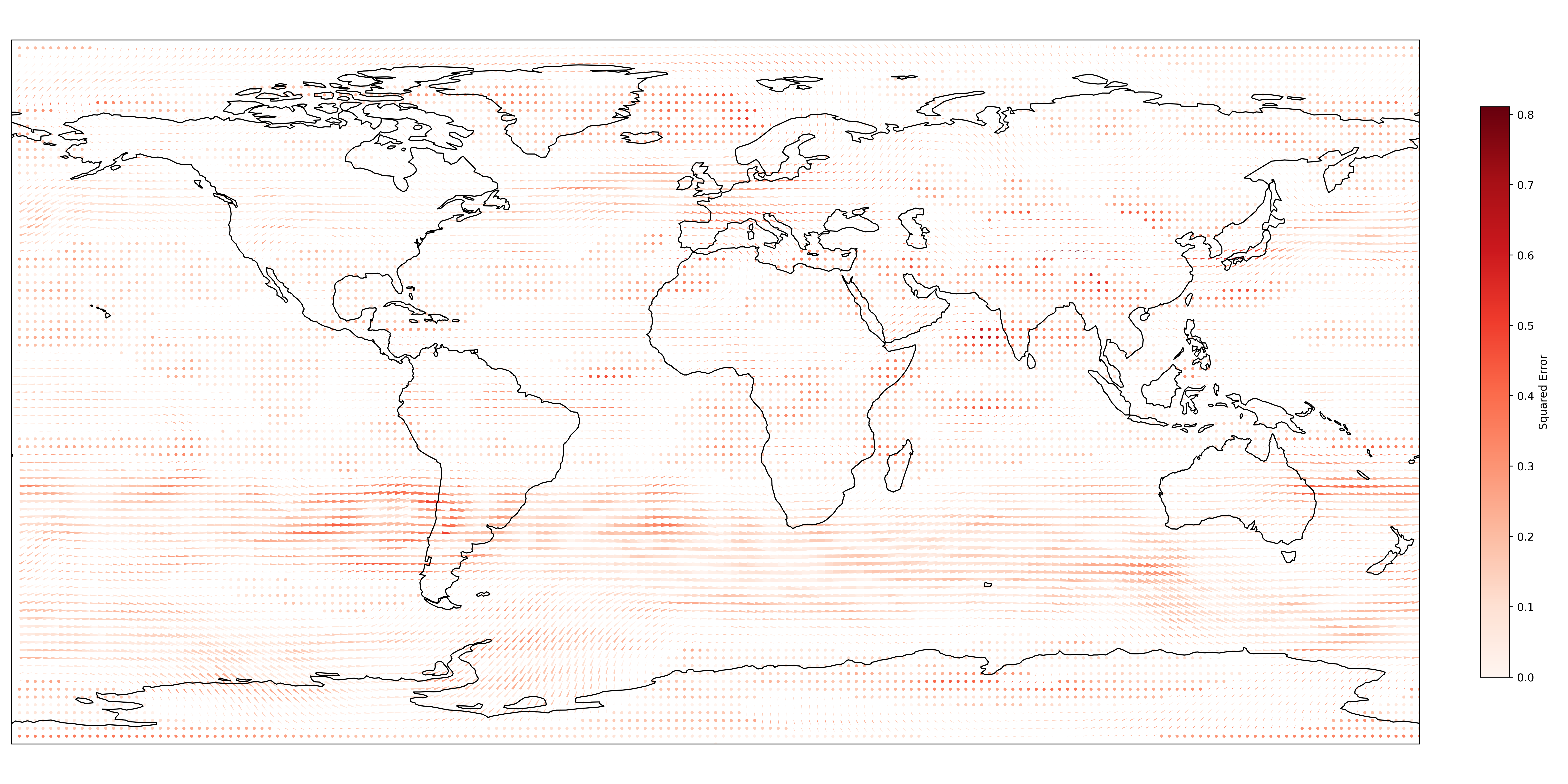}
    \caption{Squared error of downscaled monthly average wind data. Arrow length and direction are the posterior mean values, with colour giving the error magnitude.}
    \label{fig:wind3}
\end{figure}

A spherical mesh is constructed by projecting the latitude/longitude grid of prediction locations onto a sphere of radius 1 and finding the convex hull. The convex hull defines matrices $\star_0, \star_1$, and $\mathbf{d}_0$ which are used to construct the cotangent Laplacian. We can then construct the vector covariance kernel described in Section \ref{sec:VMK} from its eigenvectors $f_n$, eigenvalues $\lambda_n$, and the derivative operators defined by the mesh. This spherical mesh is an embedding of $\mathbb{S}^2$ in $\mathbb{R}^3$, and it is known via the Hairy Ball theorem that there are no harmonic vector fields on $\mathbb{S}^2$. As noted by \cite{robert-nicoudIntrinsicGaussianVector2024}, at the 500hPa pressure level winds are largely divergence-free, so we use only the curling vector field basis defined by $f^c_n = \ast(\frac{\mathbf{d}f_n}{\sqrt{\lambda_n}})^\#$, with a large $\kappa_c$ and $\nu$ = 1.5. This results in a mean squared error of 0.027 and a negative log marginal likelihood (NLL) of -652.992 with Figure \ref{fig:wind1} showing the downscaled data projected back to a latitude-longitude grid and Figure \ref{fig:wind2} showing the same data on an orthographic projection. Figure \ref{fig:wind3} shows the same vectors coloured by the squared error at each location.

\subsection{Non-Stationary High-Altitude Wind Fields}

This section considers daily mean observations of global wind fields and shows how the discrete Gaussian vector field model can be used to capture non-stationarity in the data. We take 2208 hourly observations of wind fields at the 850hPa pressure level during the 2020 Northern hemisphere winter months (November, December, January) from the ERA5 reanalysis dataset \citep{hersbachERA5HourlyData2023}, subsampled to a 10\degree \ grid on the sphere. As before, we consider only curling velocities due to the altitude of the data.

The Rossby radius of deformation \citep{gillAtmosphereOceanDynamics1982} determines the horizontal length-scales of synoptic weather systems in the upper atmosphere. It is inversely proportional to the Coriolis frequency, which varies as a constant multiple of $sin(\text{latitude})$. We would therefore expect the observed length-scales to decrease towards the poles from a maximum near the mid-latitudes. We subtract the zonal average wind velocity to remove the effect of zonal background circulation.

We consider $\kappa$ as a function of latitude, modelled using a low-rank GP built with Gaussian weights and radial basis functions. We set standard normal priors on the weights each of the basis functions. To ensure positivity, we apply a softplus transformation and set a minimum length-scale value of 1. The posterior means of the weights are inferred through MCMC.

\begin{figure}[!ht]
    \centering
    \includegraphics[width=0.75\linewidth]{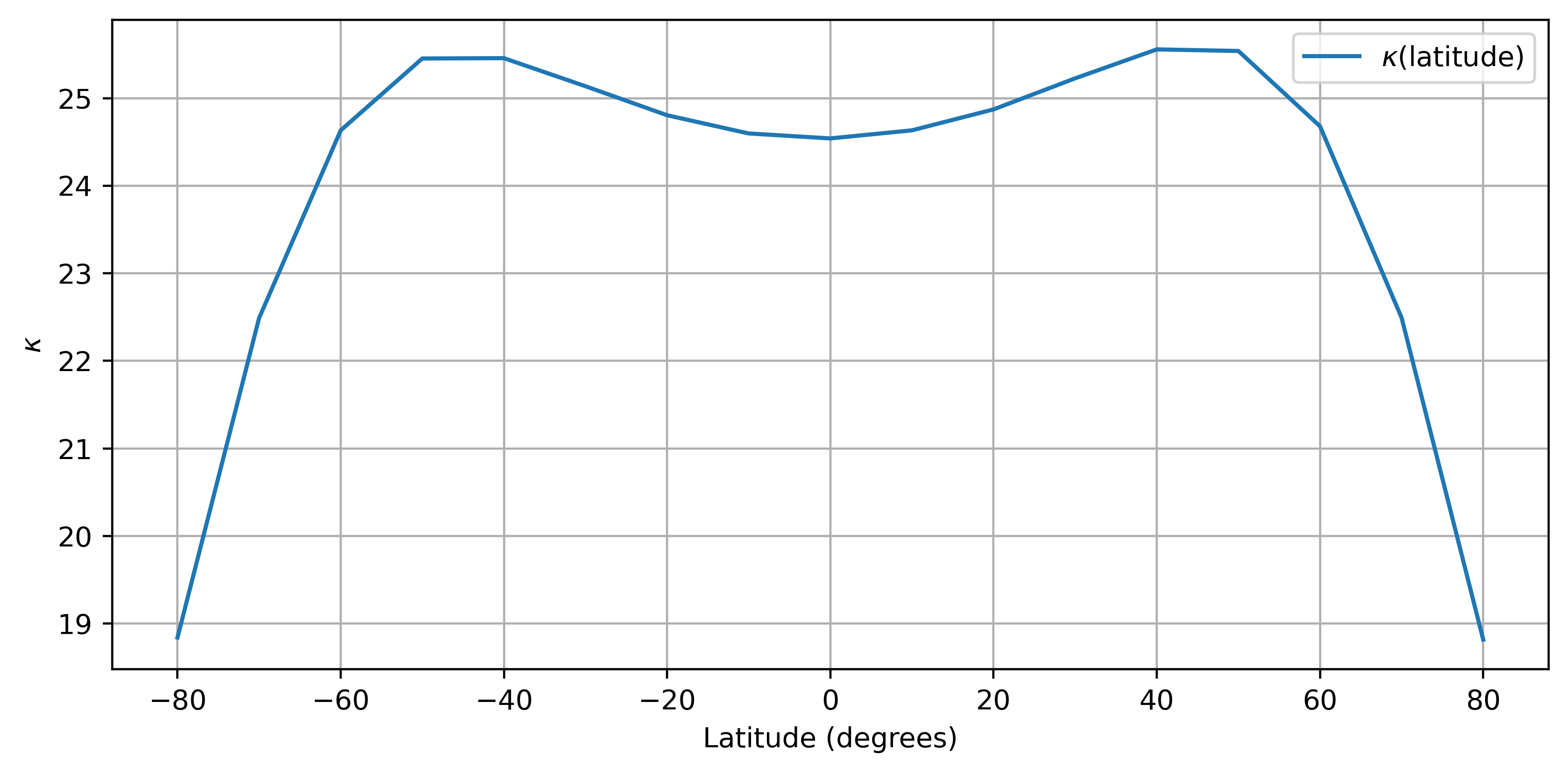}
    \caption{Low-rank basis representation of $\kappa$ as a function of latitude. 20 RBF basis kernels are used, equally spaced between -80\degree \ and 80\degree \ latitude, and the posterior means of the Gaussian weights are used to calculate a posterior mean of $\kappa$. The inferred $\kappa$ displays the expected behaviour with respect to the Rossby radius of deformation.}
    \label{fig:kappa-lat}
\end{figure}

Figure \ref{fig:kappa-lat} shows the inferred length-scale function constructed from the posterior mean of the weights. As expected, there is a smaller length-scale near the equator than the tropics, and the length-scale decreases towards the poles.

\subsection{Ocean Currents} \label{sec:ocean}

We demonstrate downscaling on a bounded domain using ocean current data in the South Atlantic and Indian Oceans around South Africa. The dataset is the Global Ocean In-Situ Near Real Time observations of ocean currents from the Copernicus Marine Data Store \citep{e.u.copernicusmarineserviceinformationcmemsGlobalOceanInsitu2018}, which collates near-surface velocity readings from drifting buoys (drifters). We use a single realisation of $u$ and $v$ near-surface velocities from 12:00:00 15th March 2025.

We enforce that the South African coastline is a no-flux boundary, but the rest of the edges of the simplicial complex are left open. Following the discussion in Section \ref{sec:harm}, we add an outer boundary away from the area of interest that is not shown. A mesh is constructed as the Delaunay triangulation of a 45 by 25 grid in the domain of interest. As before, this mesh defines the appropriate derivative operators to construct the cotangent Laplacian. To ensure the no-flux boundary, we solve the eigen-decomposition twice, first omitting the boundary points and then with the boundary points included. The first set of eigenvectors have a Dirichlet boundary condition and so are used to construct the curling basis vectors, thus ensuring that there is no flow over the boundary once the positive rotation is applied. The second set of eigenvectors are used to construct the diverging basis operators.

\begin{figure}[!ht]
    \centering
    \includegraphics[width=\linewidth]{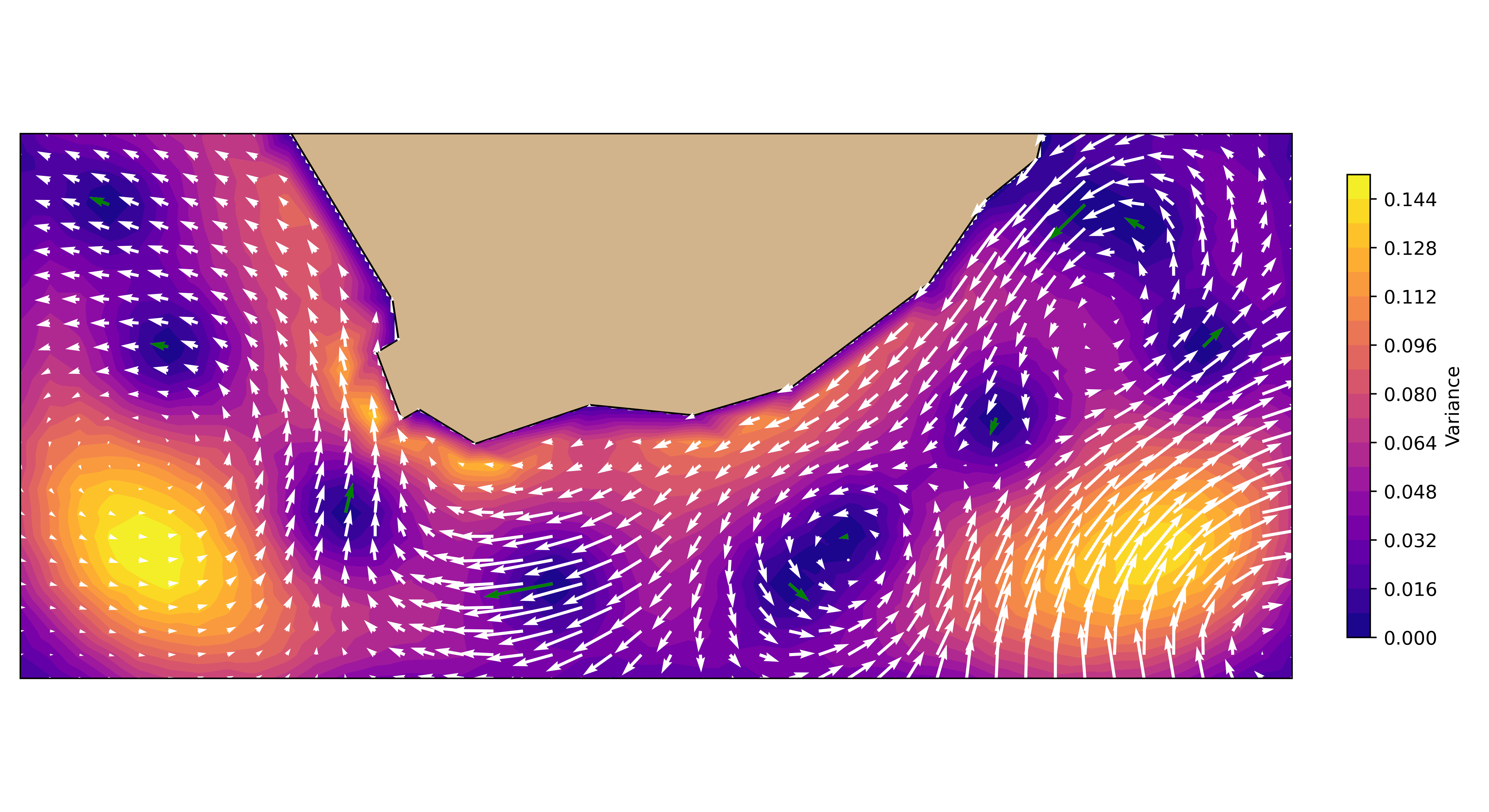}
    \caption{Downscaled ocean currents from drifter data in the South Atlantic and Indian oceans. The colour scale shows the uncertainty at each location, the green arrows are observed currents from the drifters, and the white arrows are the posterior mean field.}
    \label{fig:drifter}
\end{figure}

\begin{figure}[!ht]
    \centering
    \includegraphics[width=\linewidth]{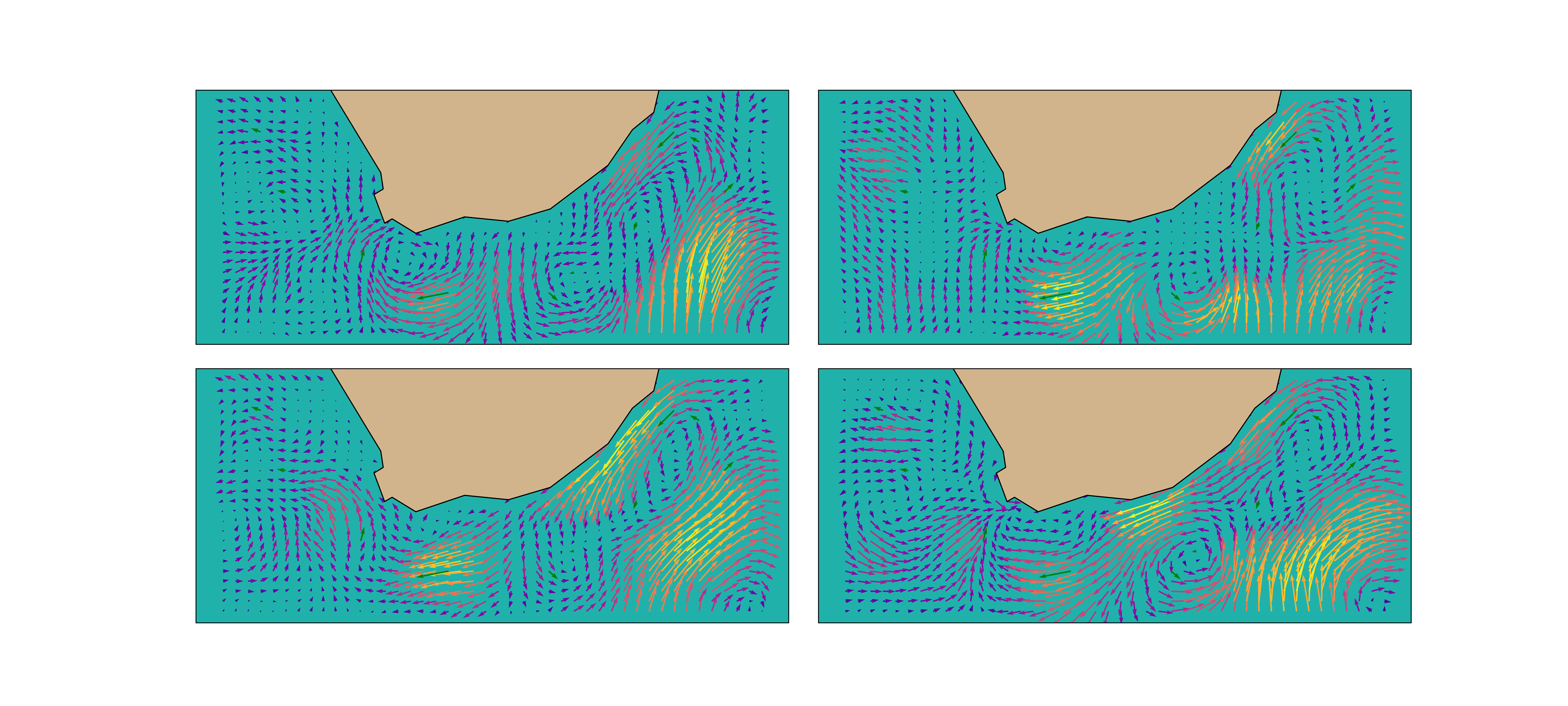}
    \caption{Four posterior samples from GP, conditioned on observed current from the drifters (green arrows). Vector magnitude is indicated by length and colour.}
    \label{fig:drift-post}
\end{figure}

We perform downscaling using the parameter values $\kappa_d= 5.0$, $\kappa_c = 50.0$, $\sigma_c = 0.5$, $\sigma_d = 2.5$, and $\nu = 1.5$, giving the results seen in Figure \ref{fig:drifter}. This figure clearly shows that the boundary conditions have been correctly applied, with no flux into the landmass of South Africa. Figure \ref{fig:drift-post} shows four different posterior samples from the GP conditioned on the observed drifter data. Despite the high variance in some regions, we can clearly see that each sample displays the same boundary conditions.

\section{Discussion}\label{sec13}

This work introduces a formulation of discrete Hodge-compositional Matérn kernels on 2-dimensional meshes that enables a flexible and powerful methodology for downscaling vector-valued data. By assuming the existence of an isomorphism to Euclidean space, it captures the metric geometry of the underlying manifold whilst also being applicable to arbitrary meshes and does not require the eigenfunctions to be known. The methodology readily extends to handle boundary conditions through a simple change to the eigenproblem, and can model data that exhibits non-stationarity through non-stationary hyperparameters.

We applied this methodology to downscaling wind fields, demonstrating how to incorporate prior knowledge about the physical processes being modelled to achieve posterior estimates with low MSE. We additionally showed how the non-stationary formulation can be used to infer non-stationary structure in data, recovering the structure of the Rossby radius of deformation from hourly wind observations. The use of boundary conditions for inferring ocean currents from sparse observations is also shown.

However, there are some possible extensions to this work in its current form. For example, it does not consider a number of useful tools when studying vector fields, such as recovering the divergence and vorticity fields. Additionally, our implementation relies on treating the Hodge star operator applied to vector fields as a rotation about the normal. This is only correct in two dimensions, and currently restricts any extension to 3-dimensional meshes. An improved formulation that allows modelling of data defined in three dimensions, such as entire atmospheric flows, is desirable.

Another possible extension concerns precision formulations of the kernels. Since our differential operators are defined with respect to the simplical complex, they exhibit significant levels of sparsity. However, in our formulation, the sparsity is lost in the eigen-decomposition, leading to dense covariance matrices. Being able to build a precision matrix for our process, which would exhibit high levels of sparsity and therefore even higher performance, is desirable. The scalar covariance matrix described in Equation \ref{eq:scalar} readily admits a precision formulation 

\begin{equation*}
     \mathbf{K}_{\nu, \kappa, \sigma^2}^{-1} = \frac{C_{\nu, \kappa}}{\sigma^2} F \ \Phi_{\nu, \kappa}(\Lambda)^{-1} \star_0^{-1} F^\top,
\end{equation*}
due to the orthogonality of the eigenvectors and since $\Lambda$ and $\star_0$ are both diagonal matrices. It is not clear if such a formulation exists for discrete Hodge-compositional Matérn covariance kernels.


\clearpage


\end{document}